\documentclass[aps,twocolumn,pre]{revtex4}  %twocolumn or preprint

\usepackage{graphicx}

\begin{document}

\title{Controllable transport mean free path of light in xerogel matrixes embedded with polystyrene spheres}
\author{Boris P.~J. Bret}
\email{bret@fisica.uminho.pt}
\author{Nuno J.~G. Couto}
%\altaffiliation{Now at Geneva}
\author{Mariana Amaro}
\altaffiliation{Now at Strathclyde University}
\author{Eduardo J. Nunes-Pereira}
\author{Michael Belsley}
\affiliation{Centro de F\'isica, Universidade do Minho, Campus de Gualtar, 4710-057 Braga, Portugal}

%\doublespacing

\date{\today}

\begin{abstract}
Xerogel matrices, made by sol-gel techniques, are embedded with polystyrene spheres to promote multiple scattering of light. Varying the concentration of the spheres inside the matrix allows one to adjust the transport mean free path of light inside the material. Coherent backscattering measurements show that a range of transport mean free paths from $90$ to $600~nm$ is easily achieved. The determination of the matrix refractive index permits a direct comparison to multiple scattering and Mie theory. Such tunable diffusive sol-gel derived samples can be further optimized as random laser materials.
\end{abstract}

\maketitle

\section{Introduction}

The study of multiple scattering of light in novel optical materials has given rise to many new physical insights in the last two decades. Without interference, or after ensemble-averaging over the configuration of the scatterers, multiple scattering can be well described as a diffusion process. In the diffusion regime, light is transported across the diffusive material as a random-walk, with the average step being the transport mean free path, $\ell$. The most interesting physics in multiple scattering nevertheless comes from interference of light. On one hand, ever-stronger disordered scattering samples have pushed towards the limit of Anderson localization of light\cite{sheng95}. On the other hand, disorder in itself, in weakly scattering systems, can lead to numerous striking phenomena, such as coherent backscattering (CBS)\cite{Albada85,Wolf85}, time-reversal mirrors\cite{thomas94,derode95}, physically uncloneable functions\cite{pappu02}, applications to urban communications\cite{moustakas00} and focussing through opaque materials\cite{vellekoop07,vellekoop08}, among others. Furthermore, disorder was found to be a very good, though peculiar, feedback mechanism for lasers, which has given rise to the so-called random lasers\cite{letokhov68,gouedard93,cao00}. In most of these phenomena, a stationary multiple-scattering material is necessary to take advantage of the disorder, and the fabrication of stationary diffusive samples with a predetermined $\ell$ is highly desirable. A promising route to produce these samples is the use of glass matrices synthesized in particular by the sol-gel technique. Indeed, sol-gel synthesized material has dramatically increased in importance over the past decades. There are many factors responsible: the technique is low cost and capable of producing uniform glasses of high purity, that are robust and possess a large internal surface area\cite{hench90,brinker90}. It has also the advantage of being a low temperature processing technique (tens of degrees Celsius) as opposed to the couple of thousands degrees Celsius used in ``normal'' glass manufacturing. This low-temperature synthesis allows the inclusion of fluorescent dyes or other optically active molecules\cite{hungerford99}. The xerogel matrices, resulting from the sol-gel process, can be used for a number of applications, for example, to study the photophysics of dye molecules and their interactions with the matrix\cite{takahashi94,nakazumi97,casalboni98}; use as solid state lasers\cite{knobbe90,rahn95,hu97,nhung03}; sensors\cite{boutin97}; non-linear effects\cite{dou97} and holograms for data storage\cite{hayashida08}. Using laser dyes in xerogel matrices where a controllable level of multiple scattering has been introduced is sure to be a fruitful model system for a variety random laser studies.

In the present article, we report on the use of the sol-gel process to produce a xerogel matrix dispersed with inert scatterers in order to obtain stationary samples with a transport mean free path that can be preselected from within a wide range of values. The process described yields matrices stable in air, with variable concentrations of scatterers within the rigid and stationary matrix. As scatterers, we used polystyrene (PS) spheres of a specific diameter. Special care was taken to obtain a wide range of PS sphere concentrations inside the matrix while avoiding any aggregation. The relevant optical characterization of the diffusive samples is presented, allowing the determination of the refractive index of the bare matrix, and the transport mean free path of the samples.

\section{Preparation of the sol-gel derived samples}
Xerogel matrices of silicon dioxide (SiO$_2$) were prepared under acidic conditions according to a method adapted from Hungerford et al\cite{hungerford06}. The sol was produced by mixing $9~ml$ of tetraethylorthosilicate (TEOS) with $3~ml$ of water containing $0.2~ml$ of $0.01~M$ HCl. This mixture was placed in an ultrasonic bath for $1~h$ before being placed in a temperature controlled enclosure (aprox. $18\ ^\circ C$) for about a month. The xerogel matrices were formed by taking $2~ml$ of the sol, mixing it with $1.5~ml$ of buffer solution and $0.5~ml$ of PS spheres solution of the desired concentration. The sol-gel derived media were produced in the form of a disc using custom disc moulds. Gelling occurred within minutes and the gels were covered and stored at ambient temperature for at least two weeks. Samples were considered ready when no discernible decrease in volume was observed. The resultant discs were about 1/8 of their initial volume with a diameter of $12.0\pm0.1~mm$ and a thickness of $1.8\pm0.1~mm$. The discs were prepared to yield samples with concentrations of polystyrene spheres between $0.16\%(V/V)$ to $3.6\%(V/V)$. By this sample preparation technique crack-free samples of good optical quality could be obtained.  TEOS (=99\%) and polystyrene spheres  ($(110\pm12)~nm$ in diameter) solution were purchased from Sigma-Aldrich and used as received.

\section{Optical characterization of the sol-gel derived samples}
\begin{figure*}[ht!]
%    \color{textcolor}
    \centering
        \includegraphics[width=120mm]{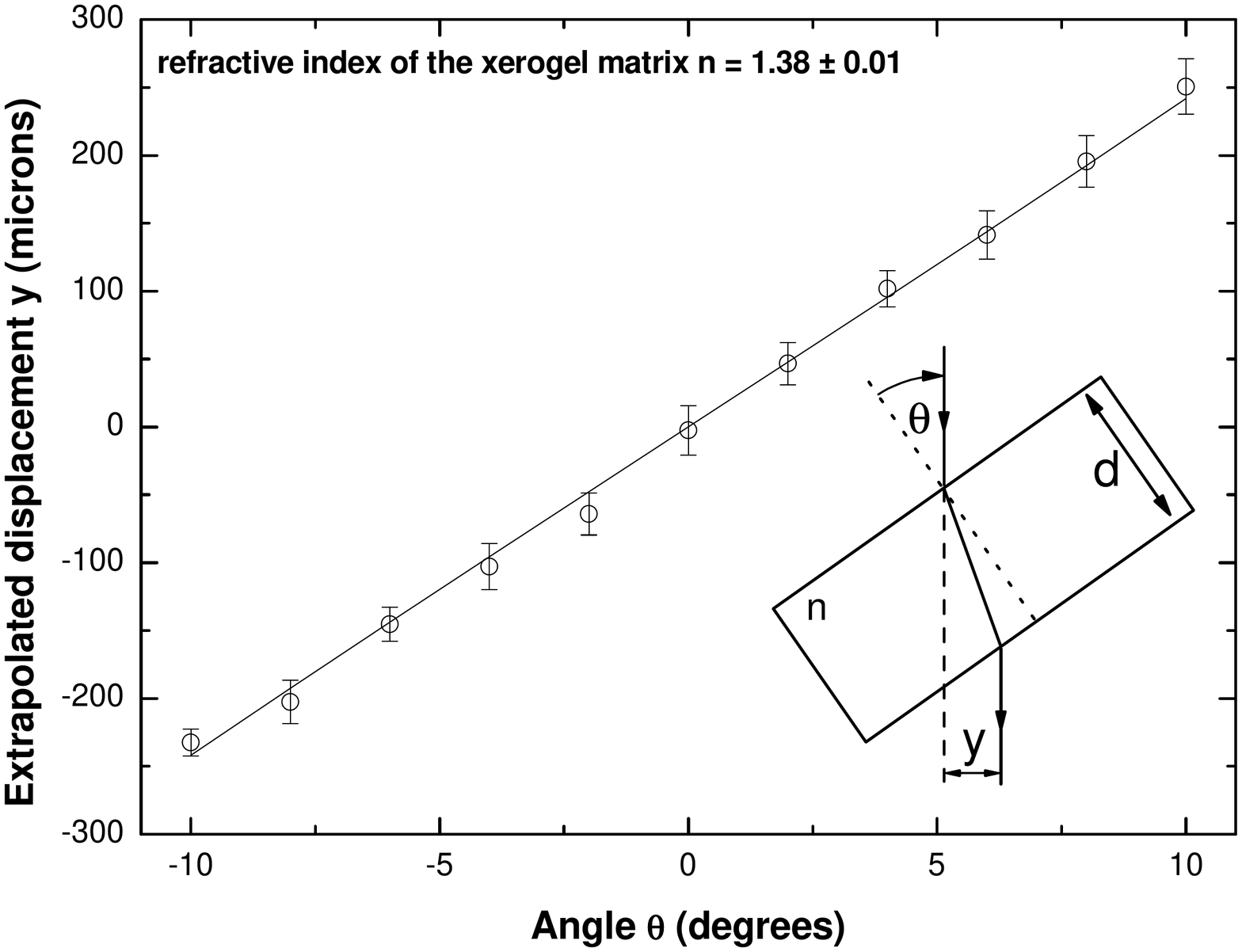}
    \caption{Determination of the refractive index of the xerogel matrix, without PS spheres. The inset presents the setup scheme. The displacement of the back face of the sample, $y$, from a laser beam impinging on the front face, is monitored as a function of incident angle $\theta$. Simple geometric optics allows the measurement of the matrix refractive index as $1.38 \pm 0.01$. Knowing the refractive index of the matrix is a key characterization of the diffusive samples.}
    \label{fig:refracsolgel}
\end{figure*}
The refractive index of the xerogel matrix was measured for a sample without any embedded PS spheres with a technique comparable to the one from Richter and Lipka\cite{richter03}. The spatial displacement of a laser beam crossing a known thickness of the sample is monitored as a function of incidence angle (see Fig.~\ref{fig:refracsolgel} ). Geometric optics leads to the following dependance:
\begin{eqnarray}y=d\left\{\sin \theta -  \cos \theta \tan \left[\arcsin\left(\frac{\sin \theta}{n}\right)\right]\right\}\label{eq:refrac} \end{eqnarray} where $y$ is the lateral displacement of the laser beam as it exits the sample, while $d$ and $n$ are respectively the thickness and refractive index of the sample (see inset of Fig.~\ref{fig:refracsolgel}). We monitored the displacement through the sample and a distance in air of a single-mode continuous-wave laser of wavelength $\lambda = 532~nm$ (Coherent Verdi) by using a beam profiler (DataRay WinCamD-UCD12). Special care was taken to ensure that the axis of rotation of the sample coincided with the point of incidence of the laser beam on the sample. The shape of the transmitted beam was found similar to the one before transmission through the sample, indicating that the surface of the sample was of good optical quality. This good surface quality is mainly a result of the smoothness of the moulds used during the sol-gel process, and the duration of the shrinking process which avoids excessive strain in the sample. Performing 5 sets of measurements of the displacement at 5 different distances between sample and detector allowed us to extrapolate back to obtain the displacement at the back face of the sample. Fig.~\ref{fig:refracsolgel} shows the dependance of this extrapolated displacement $y$ with incidence angle $\theta$. The fit of Eq.~\ref{eq:refrac} to the measurements in Fig.~\ref{fig:refracsolgel} leads to a refractive index of the xerogel matrix of $1.38 \pm 0.01$. Note that the xerogel matrix is a porous structure of SiO$_2$ with pores partially filled with remants of the sol-gel reaction (namely water and ethanol). Porosity for SiO$_2$ xerogel is known\cite{brinker90,klein98} to fall in the range 20\% to 50\%, and the fraction of non-empty pores is typically under 10\%.  Bulk SiO$_2$ has a refractive index of 1.46. Effective medium theories\cite{stroud75,braun06} predict the average refractive index of a mixture of two (or more) components provided the frontier between components is much smaller than the wavelength of light. Under those assumptions, we expect a refractive index of our xerogel matrix between 1.23 and 1.38. Our measured value for the xerogel refractive index indicates that our samples are on the lower edge of the porosity range.

\begin{figure*}[ht!]
%    \color{textcolor}
        \includegraphics[width=140mm]{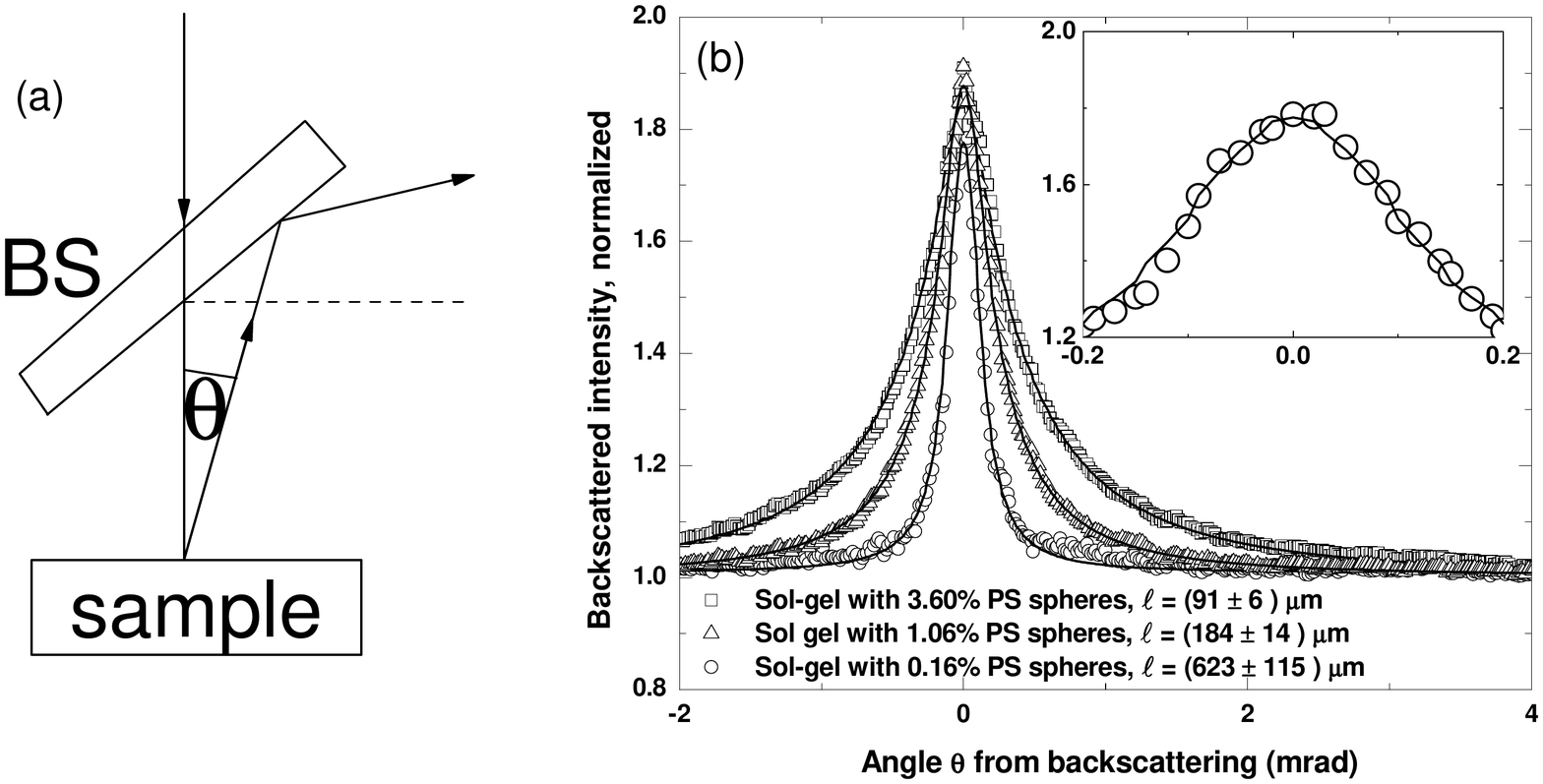}
    \caption{Determination of the transport mean free path of the sol-gel derived samples by coherent backscattering. (a) presents a schematic of the CBS setup. (b) shows the backscattered intensity vs. angle from backscattering, for three samples with different PS sphere concentration. The inset presents the detail of the top of the CBS cone for the most weakly scattering sample, rounded due to finite-thickness of the sample and setup resolution. The fit to the thickness-dependant theoretical lineshape of the CBS cone yields the transport mean free path. These samples have transport mean free path ranging from 90 to more than 600 $\mu$m. }
    \label{fig:EBSsolgel}
\end{figure*}
The main characteristic of the diffusive sample is the transport mean free path, best determined by the CBS effect. CBS is the increase, up to a factor 2 in the exact backscattering direction, of the ensemble averaged intensity reflected by a diffusive sample illuminated by a plane wave, due to the constructive interference of counter-propagating paths. The CBS can be thought of as a cone superimposed on the diffuse background, where the height of the cone itself is equal to the background, and its width is inversely proportional to the transport mean free path of the sample under study.  We use a conventional CBS setup, by illuminating the sample with a collimated laser beam and recording the sample's reflection through a beam splitter\cite{Albada85,Wolf85} (see Fig.~\ref{fig:EBSsolgel}a) on a charge-coupled device (CCD) camera. The laser used is the same as for the refractive index measurements. By using polarization optics, we record only the helicity preserving channel, which is the best channel for measuring the CBS cone\cite{wiersma95}. The overall CCD chip size and individual pixel size limits the angular range and resolution to $20~mrad$ and $0.017~mrad$ respectively. The samples under study are spun about an axis at a small angle with the incident laser beam in order to efficiently perform the necessary ensemble averaging over the stationary speckle patterns. Fig.~\ref{fig:EBSsolgel}b shows the CBS cone of three different samples, with varying PS sphere concentration. The inset of Fig.~\ref{fig:EBSsolgel}b presents the detail of the top of the CBS cone for the most weakly scattering sol-gel sample, where the rounding is due to the finite thickness of the sample\cite{mark88} and finite resolution. The finite size effect has to be considered in order to obtain good agreement between theory and experiment since the thickness of our samples ($2~mm$) is not much larger than the range of mean free paths under study and our CBS setup is sensitive to the rounding produced by this effect. The fit of the theoretical CBS lineshape\cite{mark88} to each of the CBS measurements gives the transport mean free path of each sample. The transport mean free path is found to range from $90~\mu m$ up to $600~\mu m$. This range of scattering strength is due to the varying concentrations of PS spheres embedded in the xerogel matrix. Using a less concentrated PS colloid in the sol-gel process leads to an increased mean free path. On the other hand, using a higher concentration of spheres would lead to a smaller mean free path, down to several microns. This lower limit for the method comes from the finite refractive index contrast between the xerogel matrix ($n=1.38$) and the PS spheres. Alternatively, one could use spheres with a higher refractive index (such as titanium dioxide, with a refractive index of $n=2.7$). Other more elaborate procedures allow the synthesis of diffusive samples, tunable in the very strong scattering range\cite{Wiersma97,Schuurmans99science,gomez02,storzer06,gottardo08}, around and below $\ell=1~\mu m$.

\begin{figure*}[ht!]
%    \color{textcolor}
    \centering
        \includegraphics[width=120mm]{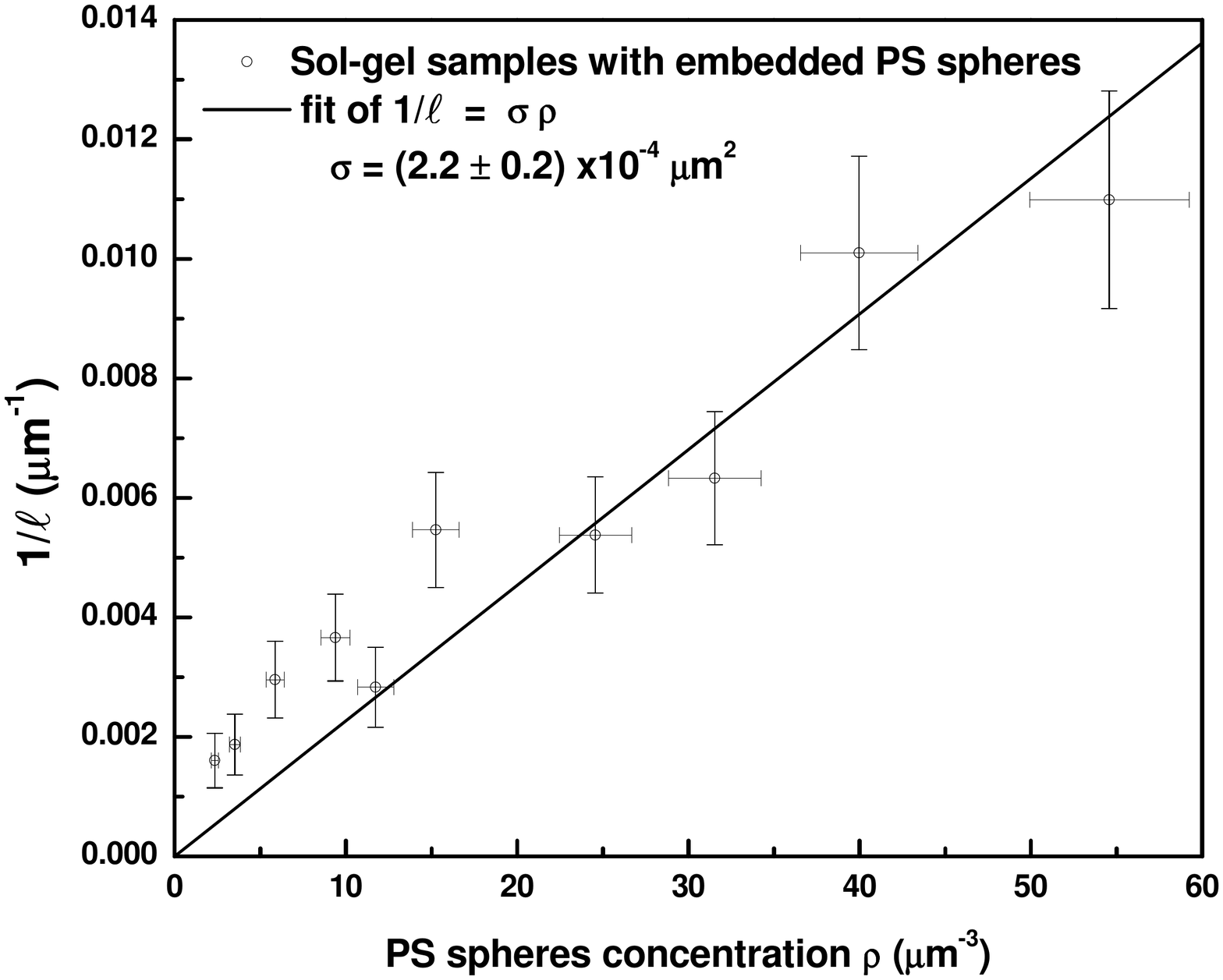}
    \caption{Determination of the scattering properties the sol-gel derived samples with embedded PS spheres. The inverse transport mean free path of the samples, measured from CBS, is shown vs. PS sphere concentration. The diffusion cross-section for the individual spheres is determined and found to be in good agreement with Mie theory predictions.}
    \label{fig:ellsolgel}
\end{figure*}
Fig.~\ref{fig:ellsolgel} presents the inverse mean free path of the samples as a function of the PS sphere concentration. The xerogel matrix being transparent to the eye, its transport mean free path is much larger than the thickness of the samples, namely $2~mm$. The matrix contribution to the scattering is therefore negligible in comparison with the scattering due to the used concentration of PS spheres, which renders the samples milky white. The transport mean free path of the sample is expected\cite{sheng95} to be inversely proportional to the concentration $\rho$ and diffusion cross-section $\sigma$ of the scatterers, as $\ell=1/(\rho \sigma)$. The slope of the fitted data in Fig.~\ref{fig:ellsolgel} directly gives the diffusion cross-section of the PS spheres, as $\sigma=(2.2\pm0.2)\times 10^{-4}~\mu m^2$. This diffusion cross-section is compatible with the predictions from Mie theory\cite{mie08,bohren83}, found to be $\sigma_{\rm Mie}=(1.4\pm0.8)\times 10^{-4}~\mu m^2$, taking into account the scattering cross section and the anisotropy factor, and considering the refractive indices of the xerogel matrix and PS spheres and the diameter of the PS spheres. The uncertainty in $\sigma_{\rm Mie}$ is dominated by the uncertainty in the PS sphere diameter ($110\pm12~nm$) quoted by the manufacturer, while the uncertainty in the refractive index of the matrix and the spheres plays a lesser role. It is therefore easy, by choosing PS spheres with the right diameter and concentration, to obtain a diffusive sample with the desired transport mean free path.

\section{Conclusion}

We have synthesized diffusive stationary samples by using a method adapted from standard sol-gel techniques, incorporating various concentrations of PS spheres before gelling. Characterization of the diffusive samples by CBS measurements confirms that we were able to fabricate samples with a transport mean free path $\ell$ ranging from 90 to $600~\mu m$. Using different concentrations of PS spheres, or spheres of different nature (size, refractive index), we expect that it is possible to vary $\ell$ down to $1~\mu m$ and up to several millimeters. Incorporating dye molecules inside the xerogel matrix\cite{hungerford99} would allow one to fabricate multiple scattering media with a preselected transport mean free path and optical gain. Such a tunable, diffusive, stationary sample with optical gain is a very promising candidate for detailed study of the physics of random lasers.

\begin{acknowledgments}This work is part of the research programme financed by the Funda\c
c\~ao para a Ciencia e a Tecnologia (FCT), through the grants SFRH/BPD/23885/2005 and PTDC/FIS/68419/2006.\end{acknowledgments}

\end{document}